\pdfoutput=1
\documentclass[aps,prl,twocolumn,groupedaddress,showpacs]{revtex4}

\usepackage{amssymb}
\usepackage{graphicx}
\begin{document}

%\preprint{}

\title{Microwave-Driven Transitions in Two Coupled Semiconductor Charge Qubits}

\author{K. D. Petersson}
\email{kdp22@cam.ac.uk}
\author{C. G. Smith}
\author{D. Anderson}
\author{P. Atkinson}
\altaffiliation{Present address: IFW Dresden, Helmholtzstra\ss{}e 20, 01069 Dresden, Germany.}
\author{G. A. C. Jones}
\author{D. A. Ritchie}
 \affiliation{Cavendish Laboratory, JJ Thomson Road, Cambridge CB3 0HE, United Kingdom}

\date{\today}

\begin{abstract}
We have studied interactions between two capacitively coupled GaAs/AlGaAs few-electron double quantum dots. Each double quantum dot defines a tunable two-level system, or qubit, in which a single excess electron occupies either the ground state of one dot or the other. Applying microwave radiation we resonantly drive transitions between states and non-invasively measure occupancy changes using proximal quantum point contact charge detectors. The level structure of the interacting two-qubit system is probed by driving it at a fixed microwave frequency whilst varying the energy detuning of both double dots. We observe additional resonant transitions consistent with a simple coupled two-qubit Hamiltonian model. 
\end{abstract}

%\keywords{}

\maketitle

Semiconductor quantum dot-based devices are fascinating systems in which it is possible to electrically define, manipulate and measure simple two-level quantum mechanical systems with a large degree of control \cite{ wiel02, fujisawa06, hanson07}. Their enormous potential to readily scale makes them attractive candidates as the building blocks for a quantum computer. One particular proposal makes use of the charge degree of freedom of a tunnel-coupled double quantum dot where the logical basis is defined by whether an electron is localized to one dot or the other \cite{fujisawa06}. This simple two-level system then forms an elementary qubit. The ability to also controllably isolate just a few electrons in a quantum dot makes it further possible to form a qubit which exploits the spin degree freedom and one such scheme is based on the singlet-triplet (S-T$_{\circ}$) spin states of a two-electron double quantum dot \cite{hanson07, petta05, taylor05}.  

To perform abitrary quantum computation operations, two-qubit gates are necessary and both charge and two-electron spin qubit schemes propose using electrostatic interactions for this purpose. Such a coupled double-dot device is also of interest as the base cell in the classical quantum dot cellular automata computational paradigm \cite{lent93, mitic06, perez-martinez07} and as a single charge qubit which is less susceptible to charge fluctuations \cite{oi05}. In this Letter we present non-invasive measurements of two capacitively coupled double quantum dots. We use microwaves to resonantly drive transitions between states and map the complex evolution of the band structure as interactions between the two qubits are turned on.
\begin{figure}
\includegraphics[scale=1]{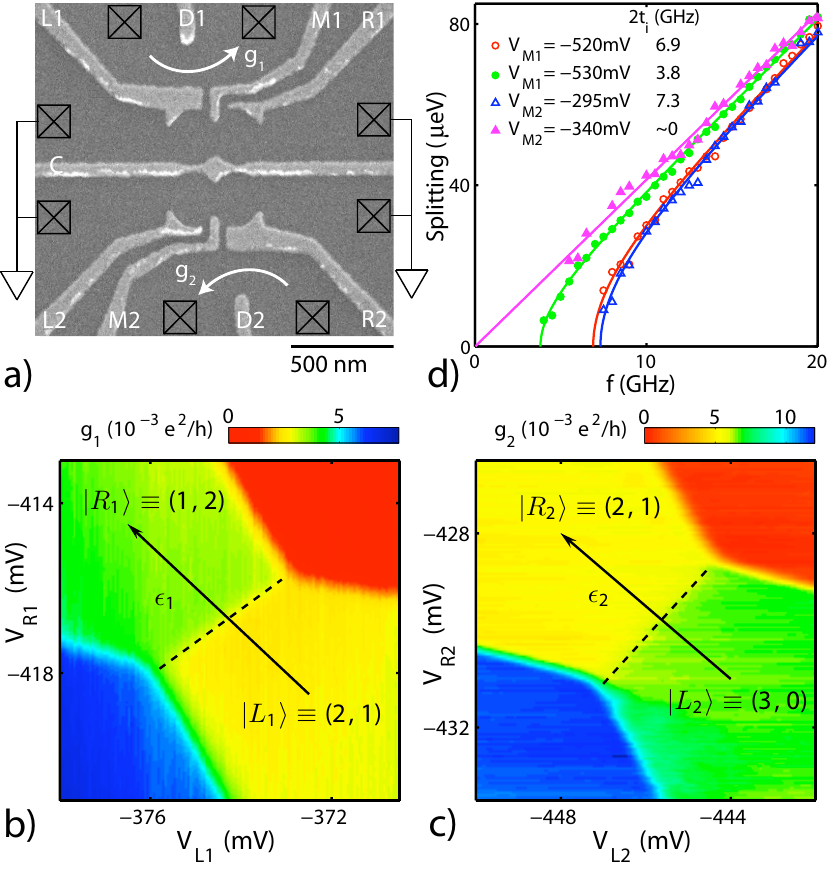}
\caption{\label{fig1} (color online) (a) Scanning electron micrograph of a control sample showing the gate pattern for the device measured. (b) Measured charge stability diagram for the top double dot showing the top QPC conductance, $g_1$, against $V_{R1}$ and $V_{L1}$. (c) Measured charge stability diagram for the bottom double dot showing $g_2$ against $V_{R2}$ and $V_{L2}$. For both (b) and (c) planes have been subtracted from the data to compensate for gate coupling.  (d) One half of the resonant microwave splitting against microwave frequency for the top (circles) and bottom (triangles) double dots. The solid lines are least-square fits to theory with the x-axis intercept giving $2t_i$. We vary $t_i$ by adjusting $V_{Mi}$. }
\end{figure}

The coupled double quantum dot device was defined using Ti/Au gates on a GaAs/AlGaAs heterostructure with a two dimensional electron gas (2DEG) 90nm below the surface. The 2DEG had a mobility of  $9.1\times10^5$ cm$^2$/Vs and a carrier concentration in the dark of $1.62\times10^{11}$  cm$^{-2}$. Our experiments were carried out in a dilution refrigerator with a base temperature of $\approx\! 30$ mK.  Figure 1(a) shows a scanning electron micrograph of a test structure for the surface gates pattern which is defined using electron beam lithography. Next to the top and bottom double dots, quantum point contacts (QPCs) were defined using gates $D1/L1$ and $D2/R2$ respectively. For all measurements, the reservoirs for each of the dots were grounded and the QPCs were operated as non-invasive charge detectors \cite{field93}. An excitation of $200$ $\mu$V was applied across the circuits of both QPCs  with the current measured using standard ac lock-in detection. The conductance through each of the QPCs was typically maintained at $\sim\! 0.13 e^2/h$ \cite{conductance}. 

Measured charge stability diagrams for both the top ($i=1$) and bottom ($i=2$) double dots are shown in Figs. 1(b) and 1(c) respectively. Electron occupancy for the left and right dots, $(m_i,n_i)$, as indicated in Figs. 1(b) and 1(c), is determined by working back from different gate voltage settings where both double dots can be completely emptied. For this sample, working with total electron numbers $m_i+n_i > 1$ afforded greater control over the tunnel coupling energies whilst permitting tunneling between the dots and reservoirs.

We concentrate on charge transitions between the localized states $|L_i\rangle$ and $|R_i\rangle$ whereby a single electron is transferred from the left dot to the right by sweeping across the dashed lines marked in Figs. 1(b) and 1(c).  Given the small number of electrons on each dot we assume that the single-particle excitation energy is significantly larger than the thermal energy. Therefore near this charge transition each double dot is well approximated by a coherent two-level system \cite{hayashi03}.  With an unpaired electron on each double dot we can neglect Pauli blockade effects \cite{johnson05}. Using pseudo-spin notation we can write the system Hamiltonian for each double dot as:
\begin{equation}H_i = \frac{1}{2}\epsilon_i\sigma^z_i+t_i\sigma^x_i\end{equation}
Here $\sigma^z_i$ and $\sigma^x_i$ are Pauli matrices, $\epsilon_i$ is the energy detuning and $t_i$ is the tunnel coupling energy. To increase $\epsilon_1$ and $\epsilon_2$ we sweep in the direction of the arrows in Figs. 1(b) and 1(c) respectively.

Microwave spectroscopy is an important technique for probing coherent interactions in many level systems such as double quantum dots \cite{oosterkamp98}. When the microwave photon energy, $hf$, matches the energy difference between the ground and an excited level, the system is driven between the two states. Petta et al. \cite{petta04} have demonstrated that changes in average charge occupancy due to microwave-induced state transitions can be detected non-invasively using a QPC.  We use this method to characterise each double dot and, in particular, determine the anticrossing gap energy, $2t_i$. For our microwave measurements attenuated coaxial lines were coupled to gates $L1$ and $R1$ via bias-tees. In all measurements presented here microwaves were applied to gate $R1$.
\begin{figure}
\includegraphics[scale=1]{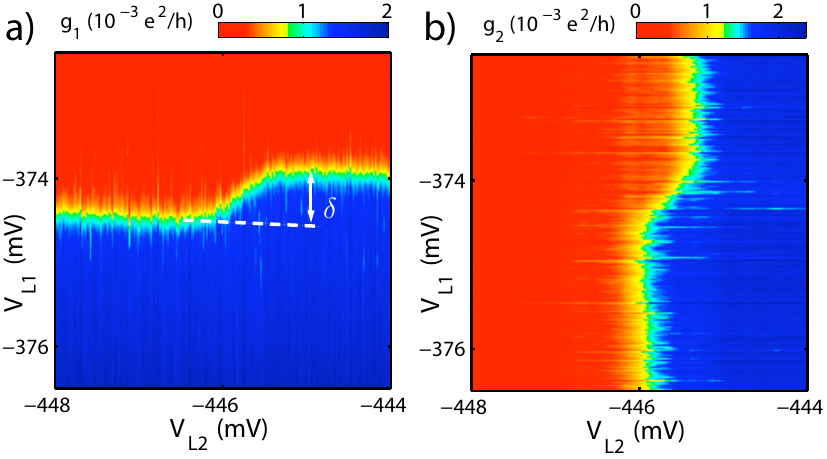}
\caption{\label{fig2} (color online) (a) and (b) plot the top and bottom charge sensing signals respectively against $V_{L1}$ and $V_{L2}$ showing the diagonal region where the two double dots interact. To account for charge detector drift in (a) we sweep $V_{L1}$ and step $V_{L2}$ whilst in (b) we sweep $V_{L2}$ and step $V_{L1}$. Each data trace has also had an offset and slope subtracted from it.}
\end{figure}

When characterising each double dot, the other double dot is biased well away from any charge degeneracies so that there are no interactions between the two halves of the device. We then measure in terms of gate voltage how the splitting between the two microwave-induced resonances, $2\Delta V_{Li}$, evolves with microwave frequency. We fit our data to $\alpha_{i} \Delta V_{Li} = \sqrt{\left(hf\right)^2-\left(2t_{i} \right)^2 }$ to obtain $2t_i$ along with the scaling factors $\alpha_i$ which convert gate voltage to energy  \cite{petta04}. Figure 1(d) maps one-half of the splitting as a function of microwave frequency for both double dots and with different voltages on gates $M1$ and $M2$. As these data show, by varying the voltage on gate $M1$ ($M2$) we can control the tunnel coupling in the top (bottom) double dot. In the case where the bottom double dot is weakly coupled ($V_{M2} = -340$ mV), $2t_2 \lesssim 2$ GHz, with the accuracy limited by thermal and inhomogeneous broadening. This measurement was also complicated by the presence of an additional splitting with a half-width of $\sim\! 3$ GHz that was observed for both double dots with sufficiently weak coupling. A similar splitting has been observed by Rushforth et al. \cite{rushforth04} who speculated it might be due to phonon absorption.

For the data presented, unless otherwise stated, we use $V_{M1} = -530$ mV and $V_{M2} = -295$ mV which yield anticrossing gap energies of $2t_1 = 3.8$ GHz and $2t_2 = 7.3$ GHz respectively.  Based on broadening of the charge transition for the top double dot we estimate an electron temperature of $\approx\!\! 110$ mK \cite{dicarlo04}. From the half-width of the microwave-induced resonances we estimate an inhomogenious charge decoherence time of $T_2^* \approx 500$ ps consistent with Petta et al. \cite{petta04}.
\begin{figure}
\includegraphics[scale=1]{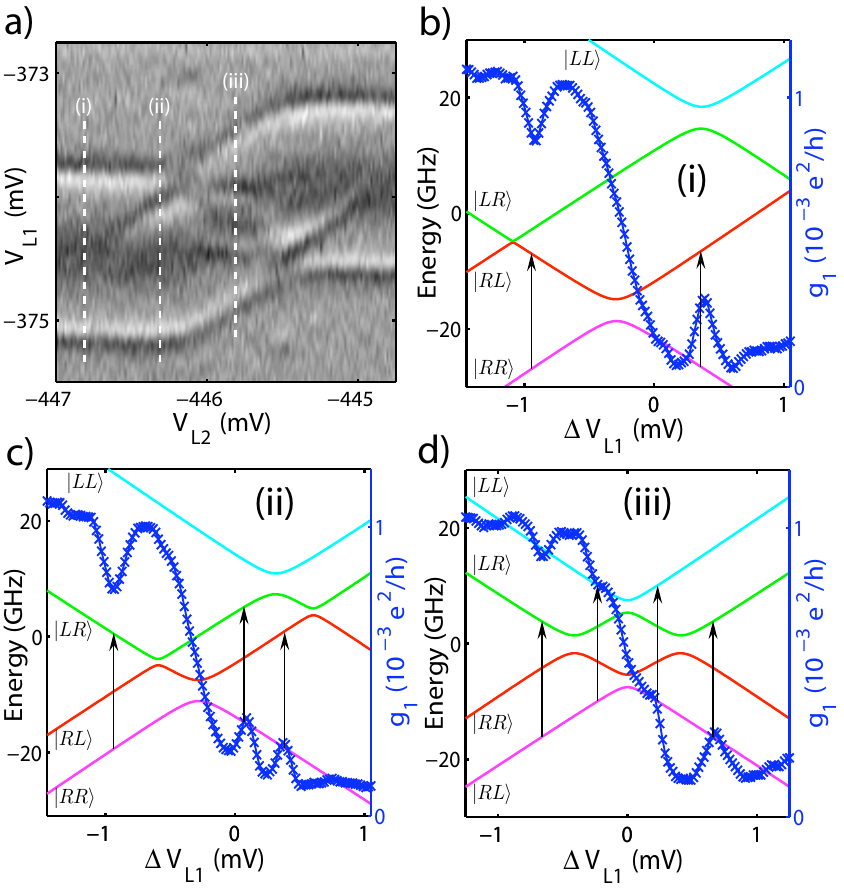}
\caption{\label{fig3} (color online) (a) We repeat the measurement in Fig. 2(a) with a $20$ GHz continuous wave signal applied. For clarity we differentiate the data with lighter regions indicating high $dg_1 / dV_{L1} $. (b)-(d) We show the detector signal traces along the dashed lines marked (i)-(iii) respectively in (a). We also plot the corrresponding calculated band-structure with arrows showing how the observed resonances coincide with $20$ GHz energy gaps between the ground state and excited states. The approximate eigenstates are indicated in the form $|LR\rangle \equiv |L_1\rangle |R_2\rangle$. }
\end{figure}

We examine interactions between the two double dots by sweeping across the charge transition for one double dot whilst stepping the other through to its charge transition point \cite{mitic06, perez-martinez07}. Figure 2(a) shows $g_1$ while sweeping across the transition for the top double dot. Figure 2(b) shows $g_2$, this time while sweeping across the transition for the bottom double dot. To account for drift in the conductance of the QPC over the long duration of these measurements ($\sim\! 4$ h), the data has been adjusted such that each sweep starts at approximately the same conductance value. A constant slope has also been subtracted from each trace. In the centre of both figures, along the diagonal feature, charge transitions coincide and both double dots are locked together by the cross-capacitance coupling energy. In this region, when one double dot is polarized by sweeping across its charge transition, polarization is simultaneously induced in the other double dot. From the shift in the charge transition position,  as indicated in Fig. 2(a), we estimate the cross-capacitance coupling energy between the two double dots to be $\delta \approx 84$  $\mu$eV ($20.3$ GHz).

In Fig. 3(a) we repeat the measurement shown in Fig. 2(a) but with a $20$ GHz continuous wave signal applied. When bottom double dot is biased away from its charge transition we observe the usual pair of resonant transitions in the top double dot detector signal [e.g., along the dashed line (i)]. However, with smaller detuning we observe additional resonances [along (ii) and (iii)] which provide strong evidence of additional levels.

To model our system we employ a total system Hamiltonian which has $\sigma^z_1\sigma^z_2$ coupling between the two qubits \cite{fujisawa06}:
\begin{equation} H = H_1+H_2+\frac{1}{4}\delta\sigma^z_1\sigma^z_2 \: . \end{equation}
We neglect $\sigma^x_1\sigma^z_2$ and $\sigma^z_1\sigma^x_2$ coupling terms on the basis that local potential changes typically have over an order of magnitude more effect on $\epsilon_i$ than $t_i$ \cite{fujisawa06}.  To convert between gate voltages and $\epsilon_i$, taking into account the cross-coupling, $\kappa_i$, between the gates, we use: 
\begin{equation} \left[ \begin{array}{cc} \epsilon_1 \\ \epsilon_2 \end{array} \right] = 
-\left[ \begin{array}{cc} \alpha_1 & \kappa_{2} \alpha_1 \\ \kappa_{1} \alpha_2 & \alpha_2 \end{array} \right] 
\left[ \begin{array}{cc} \Delta V_{L1} \\ \Delta V_{L2} \end{array} \right], \end{equation}

where $\Delta V_{Li} = V_{Li} - V_{Li}^0$ and $\Delta V_{Li} = 0$ corresponds to $\epsilon_i = 0$ (for $i=1, 2$). For our model, $\alpha_i$ ($\sim\! 0.124 - 0.134$ eV/V) and $t_i$ are measured by microwave spectroscopy of each qubit [Fig. 1(d)], whilst $\kappa_i$ ($\sim\! 0.02-0.035$), $V_{Li}^0$ and $\delta$ can be determined from charge stability plots [Fig. 2] \cite{model}.   With all model parameters we diagonalise the Hamiltonian to find the system energy levels, $E_j$, and eigenstates, $\psi_j$, $j=0, ..., 3$. In Figs. 3(b)-(d) we plot the band-structure and corresponding data sweep for three different values of $ \Delta V_{L2}$ as marked in Fig. 3(a) by the dashed lines (i)-(iii) respectively. Using our model we can explain the origin of the multiple resonances we observe in these data. For instance, in Fig. 3(d), the outer resonances occur where the photon energy matches the gap between the ground state and second excited state and charge transitions are induced in both double dots. The inner resonances correspond to transitions to the third excited state with charge transfer mainly induced in the top double dot only. 

\begin{figure}
\includegraphics[scale=1]{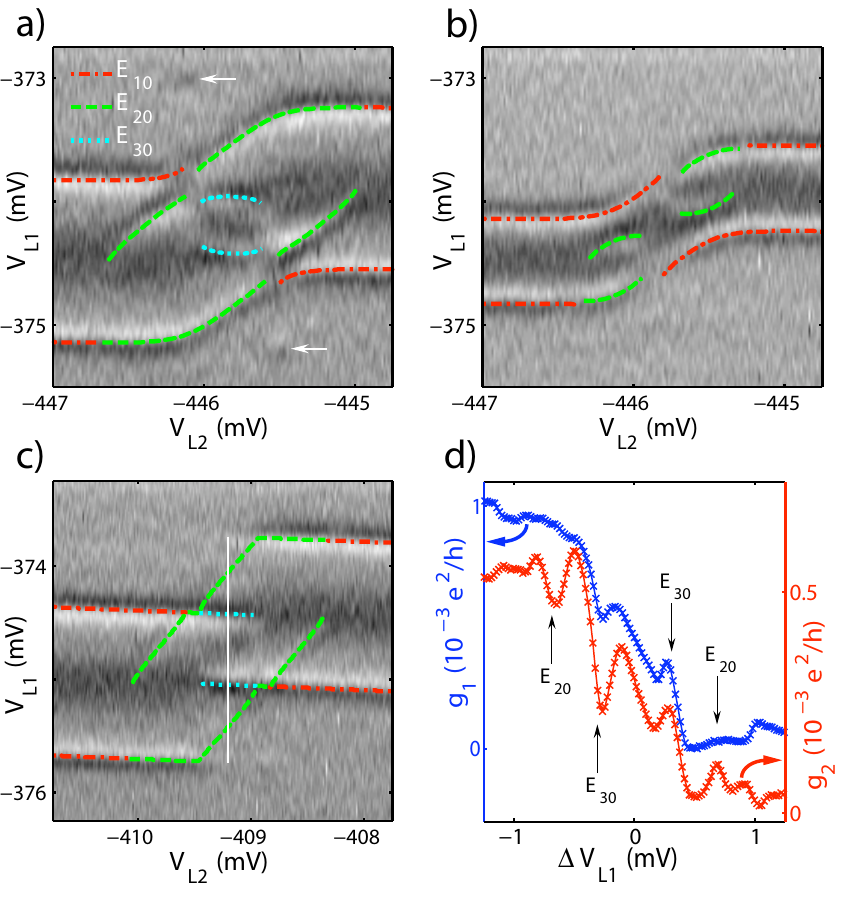}
\caption{\label{fig4} (color online) (a) Same as Fig. 3(a) but with the modelled resonances for transitions to the three excited states overlaid. The arrows indicate two photon events as described in the text. (b) We repeat (a) but with a 11 GHz continuous wave signal applied. (c) Similar to (a) but with weaker coupling in the bottom double dot. In (a)-(c) lighter regions indicate high $dg_1 / dV_{L1} $. (d) Both the top and bottom detector signals along the solid line marked in (c). The expected transition positions are indicated by the arrows. }
\end{figure}

To trace out the resonances we observe, for each value of $\Delta V_{L1}$ and $\Delta V_{L2}$ we diagonalise the Hamiltonian and map the points where $E_{j0} = E_j - E_0$ ($j=1, 2, 3$)  matches the photon energy. For the top double dot, the average difference in charge between two eigenstates $\psi_j$ and $\psi_k$ is given by:
\begin{equation}\left\langle \Delta Q_1 \right\rangle_{jk} = \left\langle \psi_j | \frac{e}{2} \sigma^z_1 |  \psi_j  \right\rangle - \left\langle \psi_k | \frac{e}{2} \sigma^z_1 |  \psi_k  \right\rangle\end{equation}
In Fig. 4(a) we overlay the resonances for $E_{j0} = 20$ GHz with the data from Fig. 3(a). As we concentrate on the top detector signal we have set the additional condition that $|\left\langle \Delta Q_1 \right\rangle_{j0}| > 0.5e$ to ensure that we only consider transitions which induce a measurable change in the average charge state of the top double dot. Noting that we have no free parameters, our model and data show good agreement. Two additional points, as marked by the arrows, are most probably due to $ E_{10} \approx E_{21} $ resulting in two-photon transitions to the second excited state via the first. 

In Fig. 4(b) we repeat Fig. 4(a) but with an 11 GHz continous wave signal applied. Around $\epsilon_2 = 0$ the expected $E_{10}$ resonances become very weak. This might occur because it is more difficult to resolve transitions at a lower frequency where effects such as inhomogenious broadening are more significant. The suppression of these resonances may also indicate a reduction in the transition rate between the two states \cite{majer05}. However, the transition rates between states depends on how exactly the microwave field couples to the device which is not known in this case.

In Fig. 4(c) we again map the top QPC detector signal against $V_{L1}$ and $V_{L2}$ with $20$ GHz applied but with reduced tunnel coupling in the bottom double dot ($V_{M2} = -340$ mV and $2t_2 \sim 0$).  We note that the expected diagonal $E_{20}$ resonances are strongly suppressed. To better understand this behavior, in Fig. 4(d) we plot both detector signal traces along $\Delta V_{L2} = 0$. The $E_{20}$ transitions can be more readily seen in the bottom detector signal. This suggests that charge dynamics are dominated by relaxation to a relatively long-lived intermediate excited state in which it is primarily the bottom double dot that is polarized.

Comparing Fig. 4(c) with Fig. 4(a), qualitatively we observe that as the tunnel coupling terms, $t_i$, are increased the resonant lines evolve from being angular to much more rounded with anticrossing-type features between them. These data agree well with our simple two qubit Hamiltonian model with $\sigma^z_1 \sigma^z_2$ coupling, demonstrating that our device behaves as a tunable four-level quantum mechanical system.

Our device design demonstrates the feasibility of coupling two few-electron double quantum dots with integral non-invasive charge detection which is key for fast read-out \cite{cassidy07, reilly07}. This is particularly relevant for two-electron spin qubits and for which, the coupling we achieve, $\delta \approx 84$ $\mu$eV, would in principle allow for a two-qubit controlled-phase operation in $\sim 25$ ps \cite{taylor05}. This time can be increased and the interaction effectively turned off by adjusting the detuning of each qubit away from $\epsilon_i = 0$. For a more realistic gate operation time of 1 ns and typical charge coherence times \cite{coherence}, the two-qubit gate error probability would be $\sim 4\%$.

\begin{acknowledgments}
We thank M. R. Buitelaar and J. R. Prance for helpful discussions and A. Corcoles for technical support. This work was supported by the EPSRC. K.D.P. acknowledges support from the Cambridge Commonwealth Trust.
\end{acknowledgments}

%\bibliography{QCA_nourl}

\begin{thebibliography}{21}
\bibitem[{mod()}]{model}
\bibinfo{note}{In practice $V_{L1}^0$ and $V_{L2}^0$ are manually adjusted when
  fitting to the data in order to take into account experimental drift.}
\bibitem[{mod()}]{coherence}
\bibinfo{note}{We assume a charge dephasing time of $T_2 = 1$ ns as measured in Ref. \cite{hayashi03} and an inelastic relaxation time of $T_1 = 15$ ns \cite{petta04}.}
\end{thebibliography}

\end{document}